# Quantum Optics of Non-Hermitian Optical Systems: Propagation of Squeezed State of Light through Dispersive non-Hermitian Optical Bilayers

Elnaz Pilehvar[1], Ehsan Amooghorban[2,3], and Mohammad Kazem Moravvej-Farshi[1*]

[1] Faculty of Electrical and Computer Engineering, Nano Plasmo-Photonic Research Group, Tarbiat Modares University, P.O. Box 14115-194, Tehran 1411713116, Iran.
[2] Faculty of Science, Department of Physics, Shahrekord University, P.O. Box 115, Shahrekord 88186-34141, Iran.
[3] Nanotechnology Research Group, Shahrekord University, Shahrekord, Iran.
* e-mail: moravvej@modares.ac.ir

We present a rigorous and quantum-consistent description of dispersive non-Hermitian optical bilayers in the framework of the canonical quantization scheme. Then we investigate the propagation of a normally incident squeezed coherent state of light through such media, particularly at a frequency for which the bilayers become parity-time ($\mathcal{PT}$) symmetric. Furthermore, to check the realization of $\mathcal{PT}$-symmetry in quantum optics, we reveal how dispersion and loss/gain-induced noises and thermal effects in such bilayers can affect quantum features of the incident light, such as squeezing and sub-Poissonian statistics. The numerical results show thermally-induced noise at room temperature has an insignificant effect on the propagation properties in these non-Hermitian bilayers. Moreover, tuning the bilayers' loss/gain strength, we show that the transmitted squeezed coherent states through the structure can retain to some extent their nonclassical characteristics, specifically for the frequencies far from the emission frequency of the gain layer. Furthermore, we demonstrate, only below a critical value of gain, quantum optical effective medium theory can correctly predict the propagation of quantized waves in non-Hermitian and $\mathcal{PT}$-symmetric bilayers.

**Keywords:** Parity-Time ($\mathcal{PT}$) Symmetry, Quantization of Electromagnetic Field, Quadrature Squeezing, Mandel Parameter

**PACS:** 95.55.Sh, 93.90.+y, 13.15.+g

## I. INTRODUCTION

Bender and co-workers [1-3] introduced a distinct class of non-Hermitian Hamiltonians that commute with the so-called parity-time ($\mathcal{PT}$) symmetric operators. In essence, a $\mathcal{PT}$-symmetric Hamiltonian commutes with the combination of the linear parity operator $\hat{\mathcal{P}}$ (i.e., $\mathbf{p} \to -\mathbf{p}$ and $\mathbf{r} \to -\mathbf{r}$) and the anti-linear time-reversal operator $\hat{\mathcal{T}}$ (i.e., $\mathbf{p} \to -\mathbf{p}, \mathbf{r} \to \mathbf{r},$ and $i \to -i$), implying that the quantum potential should satisfy the condition $V(\mathbf{r}) = V^*(-\mathbf{r})$ [3]. In other words, the real (imaginary) part of a complex potential, $V(\mathbf{r})$, is an even (odd) function of position $\mathbf{r}$. Although complex quantum potentials do not exist in nature [3], their analogs have been realized in optical systems owing to the formal equivalence between the time-dependent Schrödinger equation and the optical paraxial wave equation [4]. In this equivalence, an artificially made refractive index $n(\mathbf{r})$ plays the role of the potential, specifically in photonic waveguide array with balanced gain and loss satisfying Re $n(\mathbf{r})$ = Re $n(-\mathbf{r})$ and Im $n(\mathbf{r})$ = $-$Im $n(-\mathbf{r})$, rendering non-Hermitian systems with real eigenvalues [4-6]. These two relations together solely represent the necessary condition for the so-called *exact phase regime*. Nonetheless, beyond a critical value of gain/loss strength (i.e., the so-called *exceptional point*), the system eigenvalues become complex, no longer exhibiting the underlying $\mathcal{PT}$-symmetry. In other words, the system is in a broken $\mathcal{PT}$-symmetric phase [3].

The scattering of purely classical light from $\mathcal{PT}$-symmetric structures is a problem of both fundamental interest and practical importance that has received much attention [1-14]. These structures have been utilized to render several engaging optical effects such as optical switching [7], nonreciprocal propagation [8,9], coherent perfect absorbers [10], optical isolation [11,12], and specifically unidirectional invisibility [13,14]. These unusual functionalities, which are a direct consequence of the $\mathcal{PT}$-symmetry, are essentially classical. Since quantum light provides distinct properties, such as reduced noise and strong correlations [15] compared with classical light, hence, there is an intriguing possibility that light is treated as a stream of photons rather than classical electromagnetic waves when interacting with these structures. The optical properties of the quantum light beam propagating through lossy media are modified by absorption, dispersion, and various reflections taking place at the medium boundaries. Some of these modifications give rise to identical effects, such as shift peak and shape distortions of the transmitted pulse [16-18], in both the classical and quantum domains. However, incident light with a nonclassical nature may exhibit some alterations in the squeezing and quantum coherence that can only be described in the framework of full quantum theory [16-23]. Moreover, despite the successful implementation of $\mathcal{PT}$-symmetry in the classical systems, some controversial results are proposed in the full quantum regime, such as ultrafast states transformation and the violation of the no-signaling principle [24-25]. Given these and the apparent compensation of the losses in the $\mathcal{PT}$-symmetry systems, it would be of great interest to analyze the scattering behavior of $\mathcal{PT}$-symmetric structures in detail in a few-photons regime. This investigation may lead to fundamental and finally technological advances.

There are only a few reports devoted to exploring the effects of quantum optics on $\mathcal{PT}$-symmetric systems. Schomerus [26] has shown that a $\mathcal{PT}$-symmetric structure can become a self-sustained radiation source by a quantum noise known as microreversibility-breaking. Then, Agarwal et al. [27], using a second quantization formalism, have revealed that spontaneous generation can determine the quantum nature of the optical fields in a $\mathcal{PT}$-symmetric structure. Later on, Scheel et al. [28] investigated the effect of gain and loss on the photonic quantum states evolving in a $\mathcal{PT}$-symmetric system using the Wigner function. They found that the quantum states of light alter when propagating through the $\mathcal{PT}$-symmetric structure. Hence, one may conclude that $\mathcal{PT}$-symmetric quantum optics in a loss/gain system is unlikely. Later, Klauck et al. [29] have reported the observation of two-photon interference in an integrated, $\mathcal{PT}$-symmetric optical structure. In this work, they used Lie algebra methods to solve the quantum master equation analytically and concluded that non-local $\mathcal{PT}$-symmetric quantum mechanics act as a building block for future quantum devices. Meanwhile, Jr et al. [30] have addressed the analysis of nonclassical-light generation in a two-mode optical $\mathcal{PT}$-symmetric system. With the above background, we turn to study the propagation of squeezed coherent states through a pair of dispersive dielectric slabs forming a non-Hermitian structure, specifically when this structure is $\mathcal{PT}$-symmetric in a frequency.

In this contribution, we use the canonical quantization of the electromagnetic field [17, 18-20] in multilayer media to second quantize our system. It prepares the ground for examining the detrimental/beneficial effects of non-Hermitian and $\mathcal{PT}$-symmetric structures on incident quantum lights. The following questions naturally arise in this context have not been addressed before. (i) How and how much dispersion and loss/gain-induced and thermal noises in non-Hermitian structures affect nonclassical properties of the incident light, such as squeezing and sub-Poissonian statistics? Unidirectional invisibility is a captive phenomenon observed in a classical $\mathcal{PT}$-symmetric system. (ii) Could it be seen in a few-photons regime too? (iii) Can one correctly predict the effective behavior of such structures in quantum optics? To answer these questions, we evaluate the squeezing and the Mandel parameters of the outgoing states from the non-Hermitian bilayer structure and examine the competition between the quantum noises and the loss/gain coefficient in both exact and broken $\mathcal{PT}$-symmetric phases. These parameters can serve as appropriate measures for checking the possibility of implementation of $\mathcal{PT}$-symmetry in quantum optics.

The organization of the manuscript is as follows. In Sec. II, after introducing the suggested geometry and presenting a summary of the quantum input-output relations, the exact multilayer theory and the quantum optical effective medium theory (QOEMT) [19-21] are used to formulate the transmission and reflection amplitudes and the quantum noise fluxes emitted by the structure. Section III is devoted to simulation results analysis, demonstrating the results obtained for the eigenvalues of the scattering matrix and the corresponding transmission and reflection intensities for two types of one-dimensional non-Hermitian bilayers to find the operation regime and anisotropic transmission resonance (ATR) of $\mathcal{PT}$-symmetric bilayer. The unidirectional invisibility phenomenon introduced by Lin et al. [13] is a specific case of these ATRs. Also, we find the linear regime that the QOEMT as a "linear formalism" can accurately predict the results of the exact multilayer theory. We assume that squeezed coherent states normally incident upon the bilayer structure. Employing a balanced homodyne detector, we investigate the squeezing property of the output states. Subsequently, using a photocount detector and calculating the Mandel parameters, we study the photon-counting statistics of the output states of the bilayer structure. Finally, the paper is closed with conclusions in Sec. IV. Appendices A, B, and C provide the details of the elements of the scattering matrix, the effective noise parameters.

## II. PHYSICAL STRUCTURE AND BACKGROUND THEORY

### A. Geometry

Consider a dispersive non-Hermitian bilayer structure that is composed of a pair of gain ($g$) and loss ($l$) slabs of identical thicknesses $l_g = l_l = l$ and surrounded by vacuum (Fig. 1). For simplicity, we assume that quantum states of light, impinging at normal incidence from either side of the bilayer structure. The arrows normal to the $x$-$y$ plane together with the bosonic operators show the input and output modes. A simple schematic of the balanced homodyne detection system including, a beam splitter, two detectors, and a local oscillator (LO) field, is depicted on the right-hand side of the bilayer. Let us assume that the permittivity of the gain (loss) layer, $\varepsilon_{g(l)}$, can be described by the one-resonance Lorentz model as [31],

$$\varepsilon_{g(l)}(\omega) = \varepsilon_{bg(l)} - \frac{\alpha_{g(l)}\,\omega_{0g(l)}\,\gamma_{g(l)}}{\omega^2 - \omega_{0g(l)}^2 + i\omega\gamma_{g(l)}}, \qquad (1)$$

where $\varepsilon_{bg(l)}$ is a real number representing the background permittivity of the gain (loss) medium, $\omega$ is the input light beam frequency, $\omega_{0g(l)}$, $\gamma_{g(l)}$, and $\alpha_{g(l)}$ denote the emission

(absorption) center frequency, the corresponding linewidth, and gain (loss) coefficient. Due to the causality principle, we consider $\alpha_l > 0$ and $\gamma_l > 0$ for the loss slab. For the gain slab, we take $\alpha_g < 0$ and $\gamma_g > 0$. To guarantee the structure to be $\mathcal{PT}$-symmetric, we require an exact balance between the gain and loss of two slabs as following:

$$\mathrm{Re}\,\varepsilon_g(\omega) = \mathrm{Re}\,\varepsilon_l(\omega) \text{ and } \mathrm{Im}\,\varepsilon_g(\omega) = -\mathrm{Im}\,\varepsilon_l(\omega). \quad (2)$$

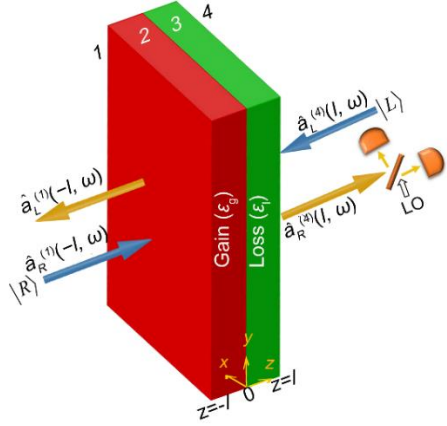

FIG. 1. (Color online) A 3D representation of a non-Hermitian bilayer structure consists of a gain and a loss slab with an identical thickness of $l$ along the $z$-direction. The arrows normal to the $x$-$y$ plane together with the bosonic operators show the input and output modes. A balanced homodyne detection is shown on the right-hand side of the structure. LO refers to the local oscillator field which is assumed to be coherent and much stronger than the input signal.

We refer the readers interested in detailed descriptions of the $\mathcal{PT}$-symmetry condition to **Appendix A.**

### B. The exact multilayer theory

Here, we use a microscopic approach to canonically quantize an electromagnetic field propagating along an optical dispersive non-Hermitian bilayer. In doing so, we start from an appropriate Lagrangian describing the electromagnetic field, a bilayer structure, and the interactions between them. We accomplish it consistently by modeling the loss(gain) layer by a reservoir composed of an infinite set of conventional(inverted) harmonic oscillators [32-33]. These harmonic oscillators, characterized by a medium field, provide the dissipation and amplification of the energy and the polarizability character of the bilayer structure and dipolarly interacting with the electric field. In this way, the bilayer structure explicitly enters the quantization scheme. By defining the canonical conjugate momentums of the system and imposing commutation relations between the dynamic variables and their conjugates, we obtain the Hamiltonian of the entire system. We also get the constitutive equations of the structure employing the Heisenberg equations for the medium field. Then, we extract the electric permeability in terms of the microscopic reservoir-field coupling function. This permeability is a complex function of frequency whose real and imaginary parts satisfy Kramers-Kronig relations. Combining the Heisenberg equations of the medium and electromagnetic fields results in a Langevin-Schrodinger equation for the electric field operator. The explicit form of the noise current density appears in terms of the initial conditions of the medium field [19-20,32]. Based on this quantization scheme, the positive frequency component of the electric field $\hat{E}$ ($z$, $t$) is given by [17-23],

$$\hat{E}_+^{(j)}(z,t) = i\int_0^\infty d\omega \sqrt{\hbar\omega/4\pi\varepsilon_0 cA} \\ \times \left[\hat{a}_R^{(j)}(z,\omega)e^{i\omega z/c} + \hat{a}_L^{(j)}(z,\omega)e^{-i\omega z/c}\right]e^{-i\omega t}. \quad (3)$$

where $A$ is the area of quantization in the $x$-$y$ plane and $\hat{a}_{R(L)}^{(j)}(z,\omega)$ represents the corresponding annihilation operator associated with the right (R) and left (L) propagating modes for the $j^{th}$ medium ($j$ = 1, 2, 3, and 4). Taking the Hermitian adjoint of Eq. (3) — i.e., $\hat{E}_-^{(j)}(z,t) = \hat{E}_+^{(j)\dagger}(z,t)$, — results in the negative frequency component of the electric field operator. Using the quantum input-output relations, the bosonic annihilation operators of the output modes on the left ($z = -l$) and right ($z = l$) boundaries (Fig. 1), $\hat{a}_L^{(1)}(-l,\omega)$ and $\hat{a}_R^{(4)}(+l,\omega)$, can be expressed in terms of their input counterparts on the opposite boundaries, —i.e., $\hat{a}_L^{(4)}(+l,\omega)$ and $\hat{a}_R^{(1)}(-l,\omega)$— and also the noise operators, $\hat{F}_{R(L)}(\omega)$ [17-23],

$$\begin{pmatrix} \hat{a}_L^{(1)}(-l,\omega) \\ \hat{a}_R^{(4)}(+l,\omega) \end{pmatrix} = \mathbb{S}\begin{pmatrix} \hat{a}_R^{(1)}(-l,\omega) \\ \hat{a}_L^{(4)}(+l,\omega) \end{pmatrix} + \begin{pmatrix} \hat{F}_L(\omega) \\ \hat{F}_R(\omega) \end{pmatrix}, \quad (4)$$

where

$$\mathbb{S} \equiv \begin{pmatrix} r_L & t \\ t & r_R \end{pmatrix} = \mathbb{A}_{22}^{-1}\begin{pmatrix} -\mathbb{A}_{21} & 1 \\ \mathbb{A}_{11}\mathbb{A}_{22} - \mathbb{A}_{12}\mathbb{A}_{21} & \mathbb{A}_{12} \end{pmatrix}, \quad (5)$$

is a 2 × 2 scattering matrix, describing the multiple transmissions and reflections within the bilayer structure. Analogous to the classical optics, $t$ represents the transmission amplitude and, $r_L$ and $r_R$ denote the reflection amplitudes through and from the bilayer binaries at $z = -l$ and $l$. The explicit forms of $\hat{F}_{R(L)}$ and the matrix elements $\mathbb{A}_{mn}$ ($m,n$ = 1,2) are given in **Appendix B.** Note that the optical input and output annihilation operators in Eq. (4) satisfy bosonic commutation relations of the form,

$$\begin{aligned} \left[\hat{a}_R^{(1)}(\omega), \hat{a}_R^{(1)\dagger}(\omega')\right] &= \left[\hat{a}_L^{(4)}(\omega), \hat{a}_L^{(4)\dagger}(\omega')\right] \\ &= \left[\hat{a}_R^{(4)}(\omega), \hat{a}_R^{(4)\dagger}(\omega')\right] \\ &= \left[\hat{a}_L^{(1)}(\omega), \hat{a}_L^{(1)\dagger}(\omega')\right] \\ &= \delta(\omega - \omega'). \end{aligned} \quad (6)$$

It is worth noting that the above quantum scattering formalism for the non-Hermitian bilayer structure can be recast into the density operator formalism by enlarging the system via introducing appropriately chosen auxiliary degrees of freedom [34]. Taking the trace on these auxiliary variables, we can get the density operator of the outgoing quantum state that describes the transformation of arbitrary input quantum states through the non-Hermitian bilayer structure. In this manner, we can successfully simulate the evolution of $\mathcal{PT}$-symmetric quantum systems in the sense of open systems [35-37].

### C. The QOEMT

In classical optics, we can describe a sub-wavelength bilayer structure of Fig. 1 by a single effective medium in which the refractive index is spatially homogeneous ($n_{\text{eff}}$). Nonetheless, one should modify the simplified model to include quantum noises via the effective noise photon distribution $N_{\text{eff}}$ in quantum optics, accompanying the classical effective refractive index $n_{\text{eff}}$ (**Appendix C**). The modified model is the so-called QOEMT [19-21]. We here briefly present the essentials needed for understating this paper. Replacing the bilayer of Fig. 1 with a single effective slab of thickness $2l$, Eq. (4) reduces to

$$\begin{pmatrix} a_L^{(1)}(-l,\omega) \\ a_R^{(4)}(+l,\omega) \end{pmatrix} = \mathbb{S}_{\text{eff}} \begin{pmatrix} a_R^{(1)}(-l,\omega) \\ a_L^{(4)}(+l,\omega) \end{pmatrix} + \begin{pmatrix} \hat{F}_{\text{eff L}}(\omega) \\ \hat{F}_{\text{eff R}}(\omega) \end{pmatrix}, \quad (7)$$

where
$$\mathbb{S}_{\text{eff}} = \begin{pmatrix} r_{\text{eff}} & t_{\text{eff}} \\ t_{\text{eff}} & r_{\text{eff}} \end{pmatrix}, \quad (8)$$

is the effective scattering matrix with the effective reflection

$$r_{\text{eff}} = \frac{e^{-2i\omega l/c}\left(n_{\text{eff}}^2 - 1\right)\left[\exp(4i\omega n_{\text{eff}} l/c) - 1\right]}{\left(n_{\text{eff}} + 1\right)^2 - \left(n_{\text{eff}} - 1\right)^2 \exp(4i\omega n_{\text{eff}} l/c)}, \quad (9a)$$

and transmission

$$t_{\text{eff}} = \frac{4n_{\text{eff}} \exp\left[2i\omega(n_{\text{eff}} - 1)l/c\right]}{\left(n_{\text{eff}} + 1\right)^2 - \left(n_{\text{eff}} - 1\right)^2 \exp(4i\omega n_{\text{eff}} l/c)}, \quad (9b)$$

amplitudes from/through the effective slab. Moreover, $\hat{F}_{\text{eff R(L)}}(\omega)$ are the effective right (left) propagating noise operators with the bosonic commutation relations given in **Appendix C**. Notice, the bosonic commutation relations (6) are still valid for the QOEMT.

## III. SIMULATION RESULTS

Considering two sets of physical parameters (Sets 1 and 2 listed in Table 1) the same as those used in [31,38] for the gain (loss) media constituting the non-Hermitian bilayer of Fig. 1, we numerically analyze the outgoing optical beam at $z = l$ for a normal incidence quantum state illuminating the left (right) bilayer boundary at $z = -l$ ( $l$ ). Moreover, we assume that the thickness of the gain (loss) layer along the $z$-direction to be $l = 10$ nm.

**Table 1.** Physical parameters for the gain and loss layers constituting the bilayer of Fig. 1, used in the simulations [31,38].

| Symbol | Definition | size Set 1 | size Set 2 | unit |
|---|---|---|---|---|
| $\varepsilon_{bl}$ | Loss layer background dielectric constant | 2 | 3.22 | – |
| $\varepsilon_{bg}$ | Gain layer background dielectric constant | 2 | 2 | – |
| $\gamma_l$ | Loss layer absorption linewidth | 67 | 140 | Trad/s |
| $\gamma_g$ | Gain layer emission linewidth | 67 | 67 | Trad/s |
| $\omega_{0l}$ | Loss layer absorption radian frequency | 1000 | 1200 | Trad/s |
| $\omega_{0g}$ | Gain layer emission radian frequency | 1000 | 1000 | Trad/s |

Notice, for Set 1(2) $\Delta\varepsilon = \varepsilon_{bl} - \varepsilon_{bg} = 0$ (1.22). Using the parameters given in Table 1, the frequency and the loss and gain coefficients that satisfy the $\mathcal{PT}$-symmetric necessary condition (2) becomes $\omega = \omega_{\mathcal{PT}} = \omega_{0g}$ (i.e., 1.885 μm) for any arbitrary value of $\alpha_l = |\alpha_g|$ for Set 1 and $\omega = \omega_{\mathcal{PT}} = 1.58\omega_{0g}$ (i.e., 1.193 μm) solely for $\alpha_l = 2$ and $|\alpha_g| = 20.86$ for Set 2 (see **Appendix A**).

### A. Linear regime

To describe the propagation of a nonclassical wave through the given bilayer, we must take care of the gain coefficient to obtain physically sound results. If the gain strength is large enough, the wave amplitude increases rapidly after a completed one round-trip and tends to saturate due to the reduction of the population inversion in the gain medium. The so-called *gain-saturation phenomenon* is a nonlinear effect associated with the round-trip parameter, characterizing a traveling wave that returns to its original position after going through two reflections at the right and left surfaces of the bilayer [18,40],

$$\eta(\omega) = \frac{\left(n_{\text{eff}}(\omega) - 1\right)^2}{\left(n_{\text{eff}}(\omega) + 1\right)^2} \exp(4i\omega n_{\text{eff}}(\omega)l/c). \quad (10)$$

Considering Eqs. (9) and (10), one can easily show that the amplitudes $r_{\text{eff}}(\omega)$ and $t_{\text{eff}}(\omega)$ have poles at frequencies for which the round-trip parameter satisfies the following conditions, simultaneously:

$$|\eta(\omega)| = 1, \quad (11a)$$

and
$$\arg[\eta(\omega)] = 0. \quad (11b)$$

Here, we focus on the linear regime of the interaction of the optical field of the incident wave with the gain slab. In particular, $|\eta(\omega)| < 1$ is a sufficient condition for obtaining analytic relations for amplitudes $r_{\text{eff}}$ and $t_{\text{eff}}$.

Using (10), we have calculated the magnitude of the round-trip parameter, $|\eta(\omega)|$, as a function of the loss coefficient, $\alpha_l$, for the two types of bilayers. The solid (dashed) curve in Fig. 2 represents the results obtained for Set 1 (2) material. A careful inspection of the data depicted by the solid curve reveals that the condition $|\eta(\omega)| < 1$ fails for $\alpha_l = |\alpha_g| \geq 147$, making analytic relations for amplitudes $r_{eff}$ and $t_{eff}$ unachievable and prediction of the exact multilayer theory by the QOEMT as a "linear formalism" inaccurate for the bilayer composed of Set 1 material. The dashed curve in Fig. 2 shows the condition $|\eta(\omega)| < 1$ holds for the entire range of given $\alpha_l$, enabling the QOEMT to predict the results of the exact multilayer theory accurately for the bilayer composed of Set 2 material.

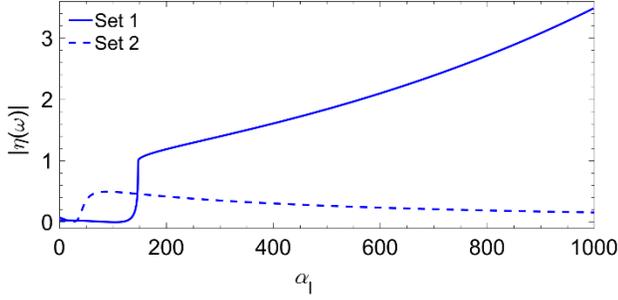

FIG. 2. (Color online) Plots of the magnitude of $\eta(\omega)$ vs. the loss coefficient $\alpha_l$ for the two types of bilayers featuring Set 1 (solid) and Set 2 (dashed).

## B. Eigenvalues and anisotropic transmission resonance

In this subsection, we turn to the eigenvalues of the scattering matrix (5) to explore the $\mathcal{PT}$-symmetry exceptional (phase breaking) points. At these points, the $\mathcal{PT}$-symmetry system transit from the exact symmetry phase with real eigenvalues to the broken symmetry phase with complex eigenvalues. Eigenvalues of the scattering matrix (5) for the bilayer of Fig. 1 are:

$$\lambda_{1(2)} = \frac{(\mathbb{A}_{12} - \mathbb{A}_{21}) \pm \sqrt{(\mathbb{A}_{12} - \mathbb{A}_{21})^2 + 4\mathbb{A}_{11}\mathbb{A}_{22}}}{2\mathbb{A}_{22}}. \quad (12)$$

For the $\mathcal{PT}$-symmetric bilayer both eigenvalues satisfy the relation $|\lambda_1 \lambda_2| = 1$ with the unimodular magnitude — i.e., $|\lambda_1| = |\lambda_2| = 1$, and the inverse moduli — i.e., $|\lambda_1| = |\lambda_2|^{-1} > 1$, characterizing the exact and the broken symmetry phases, respectively [41-42]. Using Eq. (12), we first calculate the scattering matrix eigenvalues as a function of the loss coefficient $\alpha_l$ at the $\mathcal{PT}$-symmetric frequency ($\omega_{\mathcal{PT}}/\omega_{0g} = 1$) for Set 1 (Fig. 3(a)). As can be observed from this figure, the exceptional point occurs at $\alpha_l = 890$, below (beyond) which this bilayer (Set 1) is in the exact (broken) symmetry phase regime.

Then, we obtain a similar plot for the second bilayer (Set 2) at $\omega_{\mathcal{PT}}/\omega_{0g} = 1.58$ (Fig. 3(b)). As can be seen from the inset shown in this figure, only for $\alpha_l = 2$ ($|\alpha_g| = 20.86$), at which $|\lambda_1| = |\lambda_2| = 1$ the non-Hermitian bilayer operates in the exact phase regime, recalling that for $\alpha_l \neq 2$, this bilayer (Set 2) is not a $\mathcal{PT}$-symmetric medium. Using the definitions $R_{R(L)} = |r_{R(L)}|^2$ as the reflection intensities (reflectance) from the right (left) side — i.e., at $z = +(-)l$ — and $T = |t|^2$ as the transmission intensity (transmittance) through the non-Hermitian bilayer and the symmetric property of the scattering matrix in (5), there is a generalized conservation law, $|T-1| = (R_L R_R)^{1/2}$ when the structure is $\mathcal{PT}$-symmetric with the following relationship between the right (left) reflection and transmission amplitude phases, $\phi_{R(L)}$ and $\phi_t$:

$$\phi_L = \begin{cases} \phi_R & \text{for } T < 1 \\ \phi_R - \pi = \phi_t - \pi/2 & \text{for } T > 1 \end{cases}. \quad (13)$$

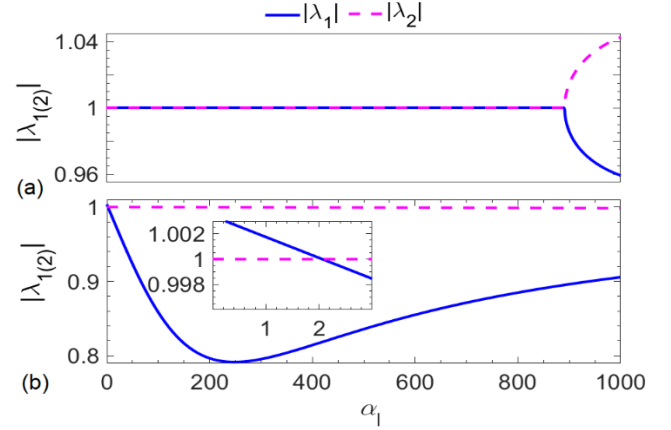

FIG. 3. (Color online) The scattering matrix eigenvalues as a function of the loss coefficient $\alpha_l$ for configuration featuring (a) Set 1, and (b) Set 2. Inset in (b) shows $|\lambda_1| = |\lambda_2| = 1$ only at $\alpha_l = 2$ ($|\alpha_g| = 20.86$).

When $T = 1$, the generalized conservation relation results in ATR — i.e., the zero reflection occurs only for the incidence from one side and not the other, while the scatterings from both sides conserve the flux [31]. For an accidental degeneracy (i.e., $R_R = R_L = R$), the flux of an incident wave conserves from both sides (i.e., $|T-1| = R$). Using Eqs. (4)-(5), we have calculated the behaviors of $T$, $R_R$, $R_L$, $\phi_t$, $\phi_R$, and $\phi_L$ versus the loss coefficient $\alpha_l$, for the bilayer structure made of materials represented by Set 1 (Fig. 4 (a)) and Set 2 (Fig. 4 (b)).

The inset in Fig. 4(a) shows the loss coefficient values ($\alpha_l \approx 24$ and 114) at which ATRs occur, and both $\phi_R$ and $\phi_L$ undergo a 180° phase change. In other words, in the range of $24 < \alpha_l < 114$ the scattering from both sides of the structure are super-unitary (i.e., $T > 1$) while in other ranges of $\alpha_l$ ($\alpha_l < 24$ and $\alpha_l > 114$) are sub-unitary (i.e., $T < 1$). Besides, there is an accidental degeneracy at $\alpha_l \sim 52$. A careful inspection of the data shown in Fig. 4(b) reveals that the transmittance is always smaller than unity, demonstrating a lack of ATR at $\alpha_l = 2$ where the conservation law satisfies.

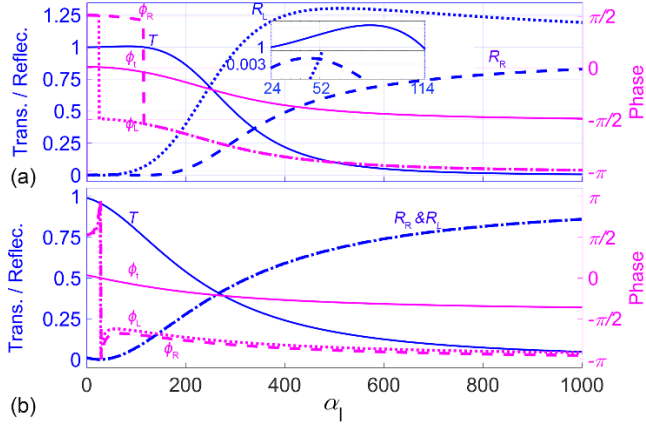

FIG. 4. (Color online) The transmission ($T$), and reflection ($R_R$ and $R_L$) intensities (Left axis) and the corresponding amplitude phases $\phi_t$, $\phi_R$, and $\phi_L$ (Right axis) vs. the loss coefficient $\alpha_l$ for the non-Hermitian bilayer composed of (a) Set 1 and (b) Set 2 materials. The inset in (a) depicts the $\alpha_l$ values at which ATRs (24 and 114) and an accidental degeneracy (52) occur.

### C. The quadrature squeezing

In this and the following subsection, we assume that a continuous-mode squeezed coherent state of light, $|R\rangle$ and quantum vacuum state, $|L\rangle$ incident from the left and right sides of the bilayer structure. The squeezed coherent state is the state with minimum uncertainty, in which the quantum fluctuations in one of the quadrature components may fall below the vacuum level at the expense of the other. We can define the squeezed coherent state of light, produced by a degenerate parametric amplifier pumped at the frequency of $2\Omega$, by the action of the squeezed operator [17-22],

$$\hat{S}\{\xi(\omega),\phi_\xi(\omega)\} = \exp\left[\int_0^{\Delta\omega} d\omega\,\xi(\omega)\hat{a}_R^{(1)}(\omega)\hat{a}_R^{(1)}(2\Omega-\omega) - h.c.\right], \quad (14)$$

and then the displacement operator,

$$\hat{D}\{\rho(\omega),\phi_\rho(\omega)\} = \exp\left[\int_0^{\Delta\omega} d\omega\,\rho(\omega)e^{-i\phi_\rho(\omega)}\hat{a}_R^{(1)\dagger}(\omega) - h.c.\right], \quad (15)$$

on the vacuum state $|0\rangle$, as following [19]:

$$|R\rangle = \hat{D}\big(\{\rho(\omega),\phi_\rho(\omega)\}\big)\hat{S}\big(\{\xi(\omega),\varphi_\xi(\omega)\}\big)|0\rangle, \quad (16)$$

where $\xi(\xi')$ and $\phi_{\xi(\xi')}(\omega)$ represent the strength and phase of the squeezed coherent state, and $\rho(\omega)$ and $\phi_\rho(\omega)$ are the amplitude and phase of the coherent component of the state $|R\rangle$. The quadrature squeezing of the scattered light on the right-hand side of the structure can be measured by a balanced homodyne detector (see Fig. 1). In this way, we mix the outgoing light from the bilayer and a strong coherent LO by a 50:50 beam splitter. Considering the detector to be ON during a sufficiently long-time-interval (i.e., a narrow-bandwidth homodyne detector), the variance of the difference between the photocurrents in the two arms of this detector at a finite temperature $\theta$ is given by:

$$\left\langle\left[\Delta\hat{E}(\omega_{LO},\phi_{LO})\right]^2\right\rangle^{out} = 1 + 2\left\langle\hat{F}_R^\dagger(\omega)\hat{F}_R(\omega)\right\rangle$$
$$+ T\left\{2\sinh^2\xi - \text{Re}(e^{2i(\phi_{LO}-\phi_t-\phi_\xi/2)}\sinh 2\xi)\right\} \quad (17)$$
$$+ R_R\left\{2\sinh^2\xi' - \text{Re}(e^{2i(\phi_{LO}-\phi_R-\phi_{\xi'}/2)}\sinh 2\xi')\right\},$$

where $\phi_{LO}$ and $\omega_{LO}$ represent the phase and frequency of the LO field and $\langle\hat{F}_R^\dagger(\omega)\hat{F}_R(\omega)\rangle$ is the average flux of the noise photons, which is given by Eq. (B7) for the exact multilayer theory and Eq. (C5) for the QOEMT. The quadrature component of the scattered light is squeezed if the variance (17) is less than unity. In our calculations, we consider $\xi = 0.2$ and $\phi_\xi = 2\phi_{LO} - 5$, with the squeezing of $\langle(\Delta\hat{E})^2\rangle^{in} = 0.926$ for the squeezed coherent state $|R\rangle$ and $\xi' = 0$ for vacuum state $|L\rangle = |0\rangle$ [17].

Making use of the exact multilayer theory and the QOEMT, we have plotted the numerical values of the variance $\langle(\Delta\hat{E})^2\rangle^{out}$ as a function of the loss coefficient $\alpha_l$ for the non-Hermitian bilayers composed of Set 1 and Set 2 materials. Figures 5(a) and 5(b) show the corresponding results at temperatures $\theta = 0$ (magenta) and 300 K (blue) when the LO frequency equals the bilayers' $\mathcal{PT}$-symmetric resonance frequencies. The solid and dashed curves show the results obtained from the exact multilayer theory and dotted and dash-dotted curves represent those of the QOEMT.

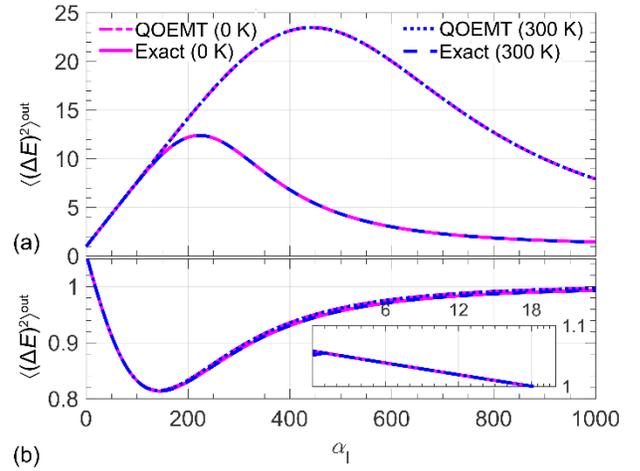

FIG. 5. (Color online) The output quadrature variance $\langle(\Delta\hat{E})^2\rangle^{out}$ as a function of the loss coefficient $\alpha_l$ for the non-Hermitian bilayers composed of (a) Set 1 and (b) Set 2 materials when $\omega_{LO}$ equals the bilayers $\mathcal{PT}$-symmetric frequencies and temperatures $\theta = 0$ (magenta) and 300 K (blue). The solid and dashed curves represent the results obtained by exact multilayer theory and dotted and dash-dotted represent those of the QOEMT. The inset in part (b) shows a zoomed-in portion in the $\alpha_l$ range where $\langle(\Delta\hat{E})^2\rangle^{out} > 1$, enhancing the visibility of the unsqueezed regime.

A careful inspection of the data shown in Fig. 5(a)

indicates that the QOEMT can accurately predict the behavior of the $\mathcal{PT}$-symmetric bilayer (Set 1) for $\alpha_l \leq 147$. Moreover, the variance $\langle(\Delta\hat{E})^2\rangle^{\text{out}}$ is always greater than unity. In other words, the incident squeezed coherent state of frequency $\omega_{\text{LO}} = \omega_{0g}$ after transmitting through the $\mathcal{PT}$-symmetric bilayer (Set 1) in the given range of $\alpha_l$ does not remain squeezed. Physically, unlike the loss slab that does not contribute to the noise flux at zero temperature, the contribution from the gain slab due to the perfect population inversion at $\omega_{0g}$ results in a maximized noise flux via the spontaneous emission of the atoms. This quantum noise contribution is significant enough to dominate the negative values of the transmission-related term (i.e., the third term on the right-hand side in Eq. (17)), resulting in $\langle(\Delta\hat{E})^2\rangle^{\text{out}} > 1$. Furthermore, this inspection reveals that the data obtained for $\theta = 0$ and 300 K coincide, indicating the insignificance of the thermal noise at room temperature for the given incident frequencies.

Moreover, we observe that the quadrature squeezing of the output state transmitted through this particular bilayer structure differs from the input state at ATR points (see Fig. (4)). This phenomenon is an apparent manifestation of broken $\mathcal{PT}$-symmetry in quantum optics for this bilayer structure. Moreover, it shows that the given bilayer in the presence of normally incident coherent squeezed light is no longer unidirectionally invisible.

Examining the data shown in Fig. 5(b), we observe excellent agreements between the data for QOEMT and the exact multilayer theory for the non-Hermitian bilayer composed of Set 2 material for $\omega_{\text{LO}}/\omega_{0g} = 1.58$ over the entire range of given $\alpha_l$ and at both temperatures. Moreover, the inset in this figurer reveals that for $\alpha_l \leq 18$ the transmitted light is unsqueezed because in the given range $\langle(\Delta\hat{E})^2\rangle^{\text{out}} > 1$. In these ranges of $\alpha_l$, where $R_R \approx R_L \approx 0$ and $T \approx 1$ (Fig. 4 (b)), the non-Hermitian bilayer behaves like a lossless/gainless slab, and hence the noise flux becomes vanishingly small. Nonetheless, unlike the case for Set 1, the transmission-related term with a small positive value contributes significantly to the squeezing, leading to $\langle(\Delta\hat{E})^2\rangle^{\text{out}} > 1$ for Set 2. For $\alpha_l > 18$, this dominant term becomes negative due to the presence of the $\alpha_l$-dependent phase, $\phi_t$, resulting in $\langle(\Delta\hat{E})^2\rangle^{\text{out}} < 1$. In other words, the incident squeezed coherent state of frequency $\omega_{\text{LO}} = 1.58\omega_{0g}$ retains its squeezed feature after transmitting through the non-Hermitian bilayer composed of Set 2 material with $\alpha_l > 18$.

For $\alpha_l \rightarrow 800$ and beyond, for either type of non-Hermitian bilayers (Set 1 and 2), as can be seen in Fig. 4, the corresponding transmittance becomes insignificant ($T \rightarrow 0$), and both reflectances ($R_R$ and $R_L$) become significantly large. Hence, both bilayers behave like mirrors, and the corresponding scattering states measured with the balanced homodyne detector approach that of the vacuum state — i.e., $\langle(\Delta\hat{E})^2\rangle^{\text{out}} \rightarrow 1$, showing contributions to the output quadrature variance from the noise- and transmission-related terms in (17) are insignificant.

Then, using the exact multilayer theory, we numerically calculated $\langle(\Delta\hat{E})^2\rangle^{\text{out}}$ versus $\omega_{\text{LO}}/\omega_{0g}$, for both non-Hermitian bilayers at $\theta = 0$ K, for a set of given $\alpha_l$ in the $\mathcal{PT}$-symmetric range (Fig. 6). Figure 6(a) shows the plots for the bilayer of Set 1, obtained for the specific values of loss coefficients whose significance was discussed in subsections III-A and III-B — i.e., $\alpha_l = 24$ (dotted), 52 (dashed), 114 (dash-dotted), 147 (solid line), and 890 (asterisks). As can be observed from Fig 3(a), each of the four $\alpha_l$ for the structures under study is in the exact symmetry phase (i.e., $\alpha_l < 890$) while for the case $\alpha_l = 890$, where the system is in a broken $\mathcal{PT}$-symmetric phase. The inset shows for each given $\alpha_l$ there is an upper limit to $\omega_{\text{LO}}/\omega_{0g}$, below which the quadrature squeezing becomes $\langle(\Delta\hat{E})^2\rangle^{\text{out}} < 1$. Those upper limits are $\omega_{\text{LO}}/\omega_{0g} = 0.73$ (dotted), 0.64 (dashed), 0.55 (dash-dotted), 0.52 (solid line), and 0.29 (asterisks). Below these upper limits, the contribution of the transmission-related term is dominant and negative enough to make $\langle(\Delta\hat{E})^2\rangle^{\text{out}} < 1$. As the incident frequency approaches to $\omega_{0g}$, the contribution from this term gradually becomes positive, and increases until $\omega_{\text{LO}} = \omega_{0g}$ along with the dominant noise flux contribution maximizes the variance $\langle(\Delta\hat{E})^2\rangle^{\text{out}}$, for all the given loss values except for $\alpha_l = 890$. The behavior of all curves near $\omega_{\text{LO}} = \omega_{0g}$ is consistent with the transmission intensities. Each plot, except that of $\alpha_l = 890$, exhibits a peak around frequency $\omega_{\text{LO}} = \omega_{0g}$. One can attribute the dip appearing at $\omega_{\text{LO}} = \omega_{0g}$ for $\alpha_l = 890$ (asterisk) to the insignificant transmission in a wide range of frequency.

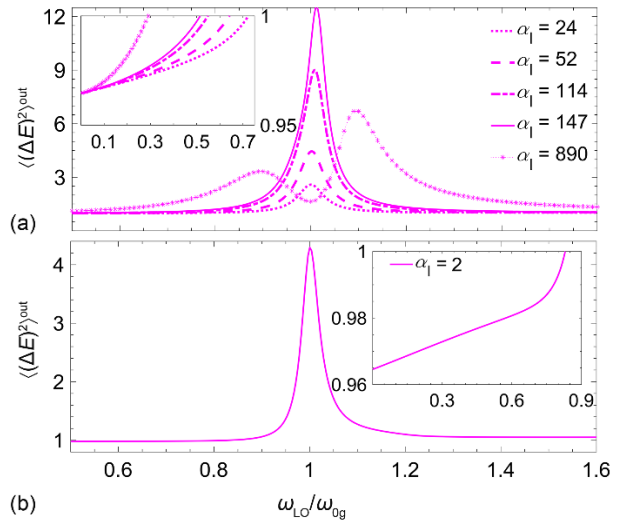

FIG. 6. (Color online) The variance $\langle(\Delta\hat{E})^2\rangle^{\text{out}}$ as a function of $\omega_{\text{LO}}/\omega_{0g}$ at $\theta = 0$ K for the non-Hermitian bilayer composed of (a) Set 1 material for $\alpha_l = 24$ (dotted), 52 (dashed) 114, (dash-dotted), 147 (solid line), and 890 (asterisks); (b) Set 2 material for $\alpha_l = 2$. The inset in each part shows a zoomed-in portion in the frequency range where $\langle(\Delta\hat{E})^2\rangle^{\text{out}} \leq 1$, enhancing the visibility of the squeezed regime.

Any further increase in frequency (i.e., $\omega_{LO} / \omega_{0g} > 1$) causes the contributions from the noise flux and the transmission-related term to drop to zero and a small positive value, respectively, eventually resulting in a variance which is slightly greater than unity.

Figure 6(b), shows the spectrum of $\langle(\Delta\hat{E})^2\rangle^{out}$ versus $\omega_{LO} / \omega_{0g}$ obtained for the only value of loss coefficient ($\alpha_l = 2$), for which the bilayer becomes $\mathcal{PT}$-symmetric at $\omega_{\mathcal{PT}} / \omega_{0g}=1.58$. The inset shows that far below the emission frequency of the gain layer (i.e., $\omega_{LO} < 0.83\omega_{0g}$), the negative contribution of the transmission-related term in Eq. (17) helps this non-Hermitian bilayer to retain the squeezing property of the incident squeezed coherent state at the output for the given $\alpha_l$. Nonetheless, $\omega_{LO} \geq 0.83\omega_{0g}$, the spectrum resembles the spectra for Set 1 in the exact-phase regime ($\alpha_l < 890$), implying similar unsqueezed output states.

### D. The Mandel parameter

To study the photon-counting statistics of the transmitted squeezed coherent state of the light through the bilayer structure, then we analyze the Mandel parameter [17]

$$Q \equiv \left\{\left\langle\left(\Delta\hat{N}\right)^2\right\rangle - \left\langle\hat{N}\right\rangle\right\}\bigg/\left\langle\hat{N}\right\rangle, \quad (18a)$$

where $\left\langle\left(\Delta\hat{N}\right)^2\right\rangle = \left\langle\hat{N}^2\right\rangle - \left\langle\hat{N}\right\rangle^2$, and

$$\hat{N} = \int_0^{T_0} dt\, \hat{a}_R^{(4)\dagger}(t)\hat{a}_R^{(4)}(t) \quad (18b)$$

is the number of photons that reach the detector over the time interval $T_0$ in region 4. The positive, zero, and negative values of the parameter ($Q$) refer to the super-Poissonian, Poissonian, and sub-Poissonian distributions.

Instead of the balanced homodyne detection in the prior subsection, here we consider a photo-count detector with a Gaussian filter function of bandwidth $\sigma_H$ and center frequency $\omega_H$, in which $\sigma_H \ll \omega_H$, to measure the Mandel parameter. Using the input-output relation (4) together with (18), after some exhausting manipulations, the Mandel parameter at finite temperature becomes

$$Q = Q_0 \Big\{(1-R_R)^2 + T^2\Big[1+\sinh^2\xi(\cosh 2\xi - 2)$$
$$+2\sigma_H\sqrt{\pi}|\rho|^2 \left(2\sinh^2\xi + \sinh 2\xi\cos(2\varphi_\rho - \varphi_\xi) - 2\right)\Big]$$
$$+2T(1-R_R)\left(\sinh^2\xi + 2\sigma_H\sqrt{\pi}|\rho|^2 - 1\right)\Big\}$$
$$\times\left\{T\left(\sinh^2\xi + 2\sigma_H\sqrt{\pi}|\rho|^2\right) + \left\langle\hat{F}_R^\dagger(\omega)\hat{F}_R(\omega')\right\rangle\right\}^{-1}$$
(19)

where $Q_0 = \sigma_H T_0 / 4\pi^{3/2}$. Following [17] we consider the Mandel parameter of $Q_i / Q_0 = -0.33$ for an incident squeezed coherent state $|R\rangle$, and presume $2\sigma_H \pi^{1/2} |\rho|^2 = 25$ and $\phi_\xi = (2\phi_\rho - \pi)$. In this manner, we prepare an incident squeezed coherent state with a sub-Poissonian statistic, demonstrating a nonclassical state. In Fig. 7, we depict the normalized Mandel parameter ($Q / Q_0$) versus $\alpha_l$, obtained for the non-Hermitian bilayer composed of (a) Set 1 and (b)

Set 2 materials for their $\mathcal{PT}$-symmetric frequency (i.e., $\omega_{\mathcal{PT}} / \omega_{0g} = 1$ and $\omega_{\mathcal{PT}} / \omega_{0g} = 1.58$) at temperatures $\theta = 0$ (magenta) and 300 K (blue). The spectra depicted by solid lines and dashes show the results obtained from the exact multilayer theory, and those displayed by dots and dash-dotted represent those of the QOEMT.

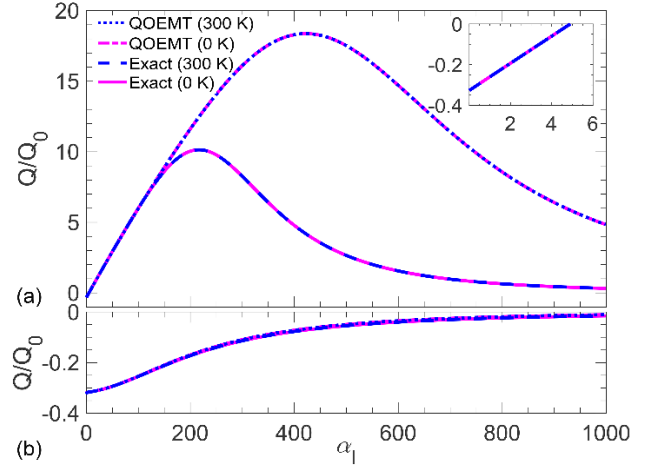

FIG. 7. (Color online) The Mandel parameter $Q / Q_0$, as a function of $\alpha_l$ for the non-Hermitian bilayers composed of (a) Set 1 and (b) Set 2 materials at $\mathcal{PT}$-symmetric frequencies at temperatures $\theta = 0$ (magenta) and 300 K (blue). The solid and dashed curves represent the results obtained by exact multilayer theory and dotted and dash-dotted curves represent those of the QOEMT. The inset in part (a) shows a zoomed-in portion in the $\alpha_l$ range where $Q / Q_0 \leq 0$, enhancing the visibility of the sub-Poissonian regime.

The inset in Fig. 7(a) shows that for $\alpha_l < 4.9$ the transmitted light is sub-Poissonian (i.e, $Q_i / Q_0 < 0$), although is not squeezed (see Fig. 5(a)). One may attribute this behavior to the insignificant reflectances ($R_R = R_L = 0$) and the transmittance of $T = 1$ (see Fig. 4(a)), in addition to the vanishing noise flux for the given range of $\alpha_l$. Knowing these facts, one can easily simplify Eq. (20), yielding $Q / Q_0 < 0$. Thus, the optical $\mathcal{PT}$-symmetric bilayer with $\alpha_l < 4.9$ can preserve sub-Poissonian statistics of incident quantum states at the output. For $\alpha_l > 4.9$, $Q / Q_0 > 0$, resulting in super-Poissonian output states. Furthermore, similar to the observations from Fig. 5 (a), data for temperatures ($\theta = 0$ and 300 K), here too coincide, implying an insignificant thermal noise effect at room temperature at incident frequency. Moreover, to our expectation, the QOEMT cannot accurately predict the behavior of the $\mathcal{PT}$-symmetric bilayer (Set 1) for $\alpha_l > 147$, (Fig. 5(a)).

Similar to observations from Fig. 5(b), data in Fig. 7(b) shows the Mandel parameter for the non-Hermitian bilayer composed of Set 2 material, obtained from both models (the exact and QOEMT) at for both temperatures ($\theta = 0$ and 300 K), coincide. Moreover, the same data reveal non-Hermitian bilayer of Set 2 can maintain the sub-Poissonian characteristic of the incident state at the output, over the entire range of given $\alpha_l$ ($Q / Q_0 < 0$). Contrary to Set 1, here we see at the $\mathcal{PT}$-symmetric frequency ($\omega / \omega_{0g} = 1.58$) that is far from the emission frequency of the gain layer, in the given set, the quantum noise is infinitesimal. In other words, some optical non-Hermitian bilayers can preserve particular nonclassical features of incident quantum states, such as sub-Poissonian statistics at their output, when the incident frequency is far from the emission frequency of the gain layer.

Finally, varying the input signal frequency, we calculated the Mandel parameter for a squeezed coherent state transmitted through the non-Hermitian bilayer composed of (a) Set 1, and (b) Set 2 with the particular loss coefficients, the same as those used in Fig. 6, at $\theta = 0$ and 300 K (Fig. 8).

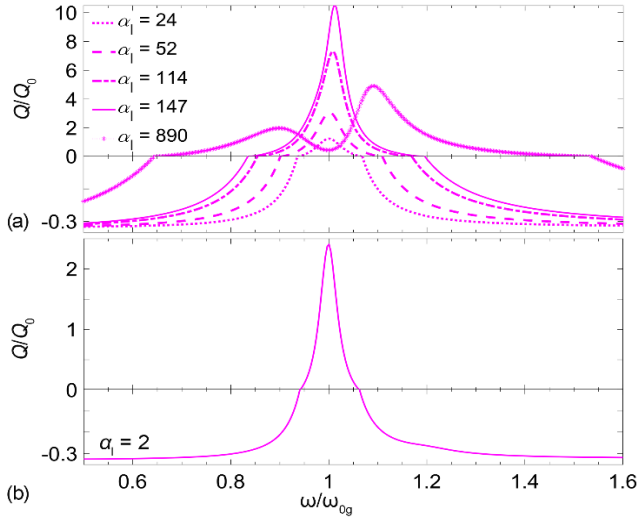

FIG. 8. (Color online) The Mandel parameter $Q / Q_0$ versus $\omega / \omega_{0g}$ for the output from the non-Hermitian bilayer composed of (a) Set 1 for $\alpha_l = 24$ (dotted), 52 (dashed), 114-(dash-dotted), 147 (solid line), and 890 (asterisks); and (b) Set 2 with $\alpha_l=2$, at $\theta = 0$. The bottom section for each part is zoomed in to enhance the visibility of the negative Mandel factor.

The dotted, dashed, dash-dotted and solid curves depict data related to $\alpha_l = 24$, 52, 114, and 147, respectively, and asterisks correspond to $\alpha_l = 890$. As can be observed from this figure, there is a frequency range over which the output light is super-Poissonian (i.e., $\omega' < \omega < \omega''$), and two other ranges over which the bilayer (Set 1) retains the sub-Poissonian feature of the incident squeezed coherent state at its output (i.e., $\omega < \omega'$ and $\omega > \omega''$). The frequencies $\omega'$ and $\omega''$ corresponding to different $\alpha_l$ values are tabulated in Table 2.

From Fig, 8(b), we can observe that in the frequency range of $0.943 < \omega / \omega_{0g} < 1.061$ the incident squeezed coherent state becomes super-Poissonian at the output of the non-Hermitian bilayer of Set 2 and remains to be sub-Poissonian out of this frequency range.

**Table 2.** The normalized frequencies ($\omega' / \omega_{0g}$ and $\omega'' / \omega_{0g}$) at which the transition of the output from the bilayer of Set 1 transits between sub- and super-Poissonian statistics, for the $\alpha_l$ range given in Fig. 8(a).

| $\alpha_l$ | 24 | 52 | 114 | 147 | 890 |
|---|---|---|---|---|---|
| $\omega'/\omega_{0g}$ | 0.936 | 0.902 | 0.854 | 0.836 | 0.645 |
| $\omega''/\omega_{0g}$ | 1.068 | 1.1 | 1.168 | 1.196 | 1.534 |

## IV. CONCLUSIONS

In the framework of the canonical quantization scheme, we have investigated the normal propagation of the squeezed coherent state of light through dispersive non-Hermitian optical bilayers, particularly at discrete sets of frequencies for which the bilayers become $\mathcal{PT}$-symmetric. In this investigation, we have calculated the output quadrature variance and the Mandel parameter at the output of each bilayer versus its loss coefficient, keeping the incident signal frequency fixed at the corresponding $\mathcal{PT}$-symmetric frequency.

Moreover, repeated the same calculations while varying the input frequency for particular sets of loss coefficients related to the $\mathcal{PT}$-symmetric regimes of the bilayers. Numerical results show, for frequencies far from the emission frequency of the gain layer, the bilayers can maintain the squeezing feature and sub-Poissonian statistics of incident squeezed coherent states to some extent at the bilayers' outputs. Here, despite the presence of the quantum noise flux, the bilayer structure composed of Set 2 makes possible the implementation of $\mathcal{PT}$-symmetry to some extent in quantum optical systems due to the loss-induced transparency phenomenon [5]. Because this structure behaves as a loss-dominated system, [Im $n^2_{eff} >0$], for a set of given $\alpha_l$ in the $\mathcal{PT}$-symmetric range, in agreement with the results in Refs. [28-29].

The numerical results also show thermally-induced noise at room temperature has an insignificant effect on the propagation properties in these non-Hermitian bilayers. Furthermore, we have shown only below a critical value of gain coefficient (linear regime), the QOEMT can correctly predict the propagation of quantized waves in non-Hermitian and $\mathcal{PT}$-symmetric bilayers.

## ACKNOWLEDGEMENTS

This work was supported by Tarbiat Modares University (TMU) under the Ph.D. Fellowship grant (IG-39703).

## APPENDIX A: $\mathcal{PT}$-SYMMETRY CONDITION

Substituting $\mathcal{PT}$-symmetry conditions (2) in (1), we can relate the gain and loss parameters:

$$|\alpha_g| = \alpha_l \frac{\omega_{0l}\gamma_l^2\left[(\omega^2-\omega_{0g}^2)^2+\gamma_g^2\omega^2\right]}{\omega_{0g}\gamma_g^2\left[(\omega^2-\omega_{0l}^2)^2+\gamma_l^2\omega^2\right]}. \quad (A1)$$

Combining (A1) with the imposed condition on the real part of permittivity in (2), after some manipulation, we get:

$$\Delta\varepsilon = \varepsilon_{bl} - \varepsilon_{bg} = \frac{\alpha_l\omega_{0l}\gamma_l\left[\omega^2-\omega_{0l}^2+\frac{\gamma_l}{\gamma_g}(\omega^2-\omega_{0g}^2)\right]}{(\omega^2-\omega_{0l}^2)^2+\gamma_l^2\omega^2}. \quad (A2)$$

For Set 1 material (i.e. $\Delta\varepsilon = 0$), from (A2) the frequency at which $\mathcal{PT}$-symmetry condition is fulfilled becomes [31]:

$$\omega = \sqrt{(\omega_{0l}^2\gamma_g + \omega_{0g}^2\gamma_l)/(\gamma_g+\gamma_l)} \equiv \omega_{\mathcal{PT}}. \quad (A3)$$

For a case in which $\omega_{0g} = \omega_{0l}$ and $\gamma_g = \gamma_l$, (A1) gives $|\alpha_g| = \alpha_l$ for all frequencies and (A2) reduces to $\omega_{\mathcal{PT}} = \omega_{0g}$. Nonetheless, for $\Delta\varepsilon \ne 0$, in general, there are at most two roots for the frequency at which the $\mathcal{PT}$-symmetry condition is satisfied. In particular for Set 2 material (i.e., $\Delta\varepsilon = 1.22$) $\omega = \omega_{\mathcal{PT}} = 1.58\omega_{0g}$ solely for $\alpha_l = 2$ and $|\alpha_g| = 20.86$.

## APPENDIX B: THE ELEMENTS OF THE SCATTERING MATRIX

The quantum noise terms originating from the absorption and amplification within the loss and gain layers can be written as [19,20],

$$\hat{F}_{R(L)}(\omega) = \mathbb{D}^{(2)}\hat{c}_{R(L)}^{(2)}(\omega) + \mathbb{D}^{(3)}\hat{c}_{R(L)}^{(3)}(\omega), \quad (B1)$$

in which

$$\mathbb{D}^{(j)} = \mathbb{A}_{22}^{-1}\begin{pmatrix} -\mathbb{B}_{21}^{(j)} & -\mathbb{B}_{22}^{(j)} \\ \mathbb{B}_{11}^{(j)}\mathbb{A}_{22}-\mathbb{A}_{12}\mathbb{B}_{21}^{(j)} & \mathbb{B}_{12}^{(j)}\mathbb{A}_{22}-\mathbb{A}_{12}\mathbb{B}_{22}^{(j)} \end{pmatrix}, \quad (B2)$$

represents a 2×2 matrix that arises from the amplification and absorption inside the gain ($j = 2$) or loss ($j = 3$) layers, with no classical analog. Notice that for the bilayer structure under study, $\mathbb{D}^{(1)} = \mathbb{D}^{(4)} = 0$. Here, the matrix $\mathbb{A}$ satisfies the following relation:

$$\mathbb{A} = \mathbb{T}^{(3)}\mathbb{R}^{(3)}\mathbb{T}^{(2)}\mathbb{R}^{(2)}\mathbb{T}^{(1)}, \quad (B3)$$

wherein $\mathbb{R}^{(j)}$ is a 2×2 diagonal matrix of elements $\mathbb{R}_{11}^{(j)} = 1/\mathbb{R}_{22}^{(j)} = \exp(-n_j''\omega l/c)$, and the elements of the matrix $\mathbb{T}^{(j)}$ are given by,

$$\mathbb{T}_{11}^{(j)} = \sqrt{\frac{n_j'}{n_{j+1}'}}\frac{n_{j+1}+n_j}{2n_j}\exp\left[i(n_j'-n_{j+1}')\omega z_j/c\right], \quad (B4a)$$

$$\mathbb{T}_{12}^{(j)} = \sqrt{\frac{n_j'}{n_{j+1}'}}\frac{n_{j+1}-n_j}{2n_j}\exp\left[-i(n_j'+n_{j+1}')\omega z_j/c\right], \quad (B4b)$$

$$\mathbb{T}_{21}^{(j)} = \mathbb{T}_{12}^{(j)}\exp\left[2i(n_j'+n_{j+1}')\omega z_j/c\right], \quad (B4c)$$

$$\mathbb{T}_{22}^{(j)} = \mathbb{T}_{11}^{(j)}\exp\left[-2i(n_j'-n_{j+1}')\omega z_j/c\right]. \quad (B4d)$$

Here, $n_j'$ ($n_j''$) is the real (imaginary) part of the refractive index $n_j$ of the medium in the $j^{th}$ region, with $n_{1(4)} = 1$, $n_{2(3)} = n_{g(l)}$, and $z_j = -l, 0,$ and $l$, for $j = 1, 2$ and $3$, respectively. For notational convenience, the frequency arguments of the above matrices are suppressed. Moreover, $\hat{c}_{R(L)}^{(2)}$ and $\hat{c}_{R(L)}^{(3)}$ represent the right (left) propagating noise operators within the gain and loss regions

$$\hat{c}_{R(L)}^{(j)}(\omega) = \pm\frac{\sqrt{2|n_j''|\omega/c}}{i}\int_{z_{j-1}}^{z_j}dz'e^{\mp in_j'\omega z'/c}f_{R(L)}^{(j)}(z',\omega), \quad (B5)$$

with the commutation relations:

$$\left[\hat{c}_{R(L)}^{(j)}(\omega),\hat{c}_{R(L)}^{(j)\dagger}(\omega')\right] = 2e^{\mp\frac{n_j''\omega l}{c}}\sinh\left(\frac{n_j''\omega l}{c}\right)\delta(\omega-\omega'), \quad (B6a)$$

$$\left[\hat{c}_{R(L)}^{(j)}(\omega),\hat{c}_{L(R)}^{(j)\dagger}(\omega')\right] = -2\frac{n_j''}{n_j'}e^{\mp\frac{n_j'\omega l}{c}}\sinh\left(\frac{n_j'\omega l}{c}\right)\delta(\omega-\omega'). \quad (B6b)$$

Here, $f_{R(L)}^{(j)}$ are the Fourier transform of the bosonic operators in the $j^{th}$ layer. Using the input-output relation in (3) and (B2)-(B6), one can find that the quantum noise operators in (B1) have the following noise flux correlation:

$$\langle\hat{F}_R^\dagger(\omega)\hat{F}_R(\omega')\rangle = \delta(\omega-\omega')$$
$$\times 2\sum_{j=2,3}\Bigg\{-\sinh(n_j''\omega l/c)\left(\left|\mathbb{D}_{21}^{(j)}\right|^2 e^{-n_j''\omega l/c}+\left|\mathbb{D}_{22}^{(j)}\right|^2 e^{n_j''\omega l/c}\right)$$
$$-\frac{|n_j''|}{n_j'}\sin(n_j'\omega l/c)\left(\mathbb{D}_{21}^{(j)*}\mathbb{D}_{22}^{(j)}e^{in'\omega(z_j+z_{j-1})/c}+\mathbb{D}_{21}^{(j)}\mathbb{D}_{22}^{(j)*}e^{-in'\omega(z_j+z_{j-1})/c}\right)\Bigg\}$$
$$\times\Big\{N_{th}(\omega,\theta)\Theta[\mathrm{Im}\,\varepsilon_j(\omega)]+(N_{th}(\omega,|\theta|)+1)\Theta[-\mathrm{Im}\,\varepsilon_j(\omega)]\Big\}. \quad (B7)$$

where $\theta$ and $\Theta$ represent the temperature and the step function and $N_{th}(\omega,\theta) = [\exp(\hbar\omega/k_B\theta)-1]^{-1}$ is the mean number of thermal photons emitted from the bilayer structure, in which $\hbar$ and $k_B$ are the reduced Planck's and Boltzmann's constants. Notice that Eq. (B7) is valid for the normal incidence at any given temperature.

## APPENDIX C: THE EFFECTIVE NOISE PARAMETERS

The right ($R$) and left ($L$) components of the quantum effective noise $\hat{F}_{\mathrm{eff}}(\omega)$ in the quantum input-output relation (7) are given by [19-20]:

$$F_{\mathrm{eff}\,L}(\omega) = \frac{-2i\sqrt{2n_{\mathrm{eff}}'n_{\mathrm{eff}}''}}{(n_{\mathrm{eff}}+1)^2-(n_{\mathrm{eff}}-1)^2\exp[4in_{\mathrm{eff}}\omega l/c]}$$
$$\times\Bigg\{(n_{\mathrm{eff}}-1)e^{4in_{\mathrm{eff}}\omega l/c}\times\int_{-l}^{l}dz'e^{-in_{\mathrm{eff}}\omega z'/c}f_{\mathrm{eff}\,R}(z',\omega)$$
$$+(n_{\mathrm{eff}}+1)\int_{-l}^{l}dz'e^{in_{\mathrm{eff}}\omega z'/c}f_{\mathrm{eff}\,L}(z',\omega)\Bigg\}, \quad (C1)$$

$$F_{\text{eff }R}(\omega) = \frac{-2i\sqrt{2n'_{\text{eff}}n''_{\text{eff}}}\exp[2i(n_{\text{eff}}-1)l]}{(n_{\text{eff}}+1)^2 - (n_{\text{eff}}-1)^2\exp[4i\omega n_{\text{eff}}l/c]}$$

$$\times \Big((n_{\text{eff}}+1)\times\int_{-l}^{l}dz'e^{-in_{\text{eff}}\omega z'/c}f_{\text{eff }R}(z',\omega) \quad (C2)$$

$$+(n_{\text{eff}}-1)\int_{-l}^{l}dz'e^{in_{\text{eff}}\omega z'/c}f_{\text{eff }L}(z',\omega)\Big).$$

with the commutation relations:

$$\left[\hat{F}_{\text{eff }R(L)}(\omega),\hat{F}^{\dagger}_{\text{eff }R(L)}(\omega')\right] = \left(1-|r_{\text{eff}}|^2-|t_{\text{eff}}|^2\right)\delta(\omega-\omega'), \quad (C3)$$

$$\left[\hat{F}_{\text{eff }R(L)}(\omega),\hat{F}^{\dagger}_{\text{eff }L(R)}(\omega')\right] = -\left(r_{\text{eff}}t^*_{\text{eff}} + r^*_{\text{eff}}t_{\text{eff}}\right)\delta(\omega-\omega'). \quad (C4)$$

Furthermore, by using Eq. (C1)-(C4), the flux of noise photons emitted by the effective slab at finite temperature $\theta$ can be written as,

$$\left\langle \hat{F}^{\dagger}_{\text{eff }R(L)}(\omega)\hat{F}_{\text{eff }R(L)}(\omega')\right\rangle = \left(1-|r_{\text{eff}}|^2-|t_{\text{eff}}|^2\right)\delta(\omega-\omega')$$

$$\times\Big\{N_{\text{eff}}(\omega,\theta)\Theta\left(\text{Im }n^2_{\text{eff}}(\omega)\right)$$

$$-\left(N_{\text{eff}}(\omega,|\theta|)+1\right)\Theta\left(-\text{Im }n^2_{\text{eff}}(\omega)\right)\Big\}, \quad (C5)$$

where

$$N_{\text{eff}}(\omega,\theta) = -\frac{1}{2} + \frac{1}{2}\sum_{j=g,l}p_j\frac{\left|\text{Im }n^2_j(\omega)\right|}{\left|\text{Im }n^2_{\text{eff}}(\omega)\right|}\left(2N_{\text{th}}(\omega,|\theta|)+1\right), \quad (C6)$$

is the effective noise photon distribution. Here, $p_j = l_j/(l_g + l_l)$, $(j = g, l)$ are the volume fractions of the layers and equals $1/2$ for the bilayer structure of Fig.1. Moreover, the effective refractive index $n_{\text{eff}}$ can be obtained from the following Bloch dispersion relation in the long-wavelength limit,

$$\cos(2n_{\text{eff}}\omega l/c) = \cos(n_g\omega l/c)\cos(n_l\omega l/c)$$

$$-\frac{1}{2}\left(\frac{n_g}{n_l}+\frac{n_l}{n_g}\right)\sin(n_g\omega l/c)\sin(n_l\omega l/c). \quad (C7)$$

Here, $n_g$ and $n_l$ are the refractive indices of the gain and loss layers that support the Bloch waves.